# Novel Relay Selection Protocol for Cooperative Networks

## Authors:

Muhammad Asam,
asim2k994@gmail.com
PhD Scholar

Zeeshan Haider,
luckier19@gmail.com
PhD Scholar

Tauseef Jamal,
**tauseef.jamal@lx.it.pt**
Sr. Researcher IT Lisboa

Kashif Ghuman

imkashifghumman@gmail.com

Aleena Ajaz

aleena.ajazsh@gmail.com

*Abstract—* **Extensive research has been done to achieve better throughput and reliability in wireless networks, with focus on multiple-input multiple-output systems. Recently, cooperative networking techniques have been investigated to increase the performance of low-cost wireless systems by using the diversity created by different single antenna devices. However, cooperative networking requires a medium access control layer able to handle source-relay-destination communications. Wireless cooperative relaying poses several challenges, being the most important one related to the relay selection mechanism, especially in the presence of mobile nodes. This position paper aims to describe our findings towards development of an efficient relay selection algorithm.**

*Index Terms—* **Cooperative relaying, opportunistic relaying, channel estimation**

## I. INTRODUCTION

WIRELESS networking provides easy connectivity and fast deployment, but still presents low performance level. The major limitation of wireless networks comes from the shared medium and the unstable wireless channel. Chan- nel conditions in wireless networks are subjected to fading variations (c.f. 1), including interference which can affect both throughput and reliability.

Transmitting independent copies of the signal generates diversity trying to mitigate the effects of fading. For instance, spatial diversity is created when the different copies of the same signal are transmitted from different locations, while time diversity is created when different copies of the same sig- nal are transmitted by the same device in different transmission opportunities. In both cases, the receiver gets independently faded versions of the signal. In what concerns the space domain, cooperative relaying techniques are able to generate the required diversity. However, at the link layer, the *Medium Access Control* (MAC) would have to handle more than one- hop communication, being distributed and cooperative in a multipoint-to-multipoint communication.

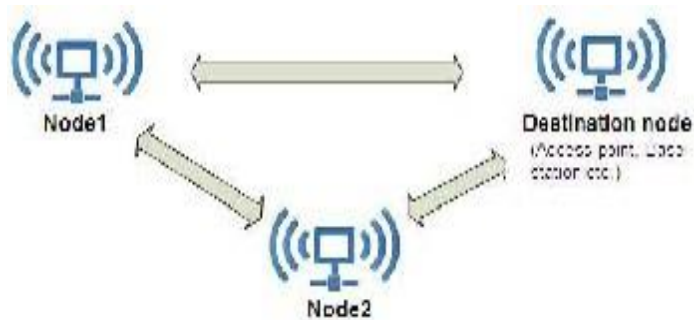

Fig. 1. Fading paths

One of the most important mechanisms to be included in a cooperative relaying mechanism is the process of selecting one or more relays. Relay selection must be efficient since relaying requires more reception events than the direct source to destination transmission, and so may lead to lower energy efficiency. However, the great majority of research on cooperative networking focused on the investigation of exploiting wireless diversity, and few approaches have been proposed to analyze the impact and requirements of cooperative relaying above the physical layer.

## II. STATE OF THE ART

The performance of cooperative relaying schemes mainly depends on an efficient relay selection, since it can drasti- cally improve the performance of data transmissions. Such mechanisms should be distributed in order to be applied to dynamic multi-hop networks. However, distributed relay selection requires the exchange of messages that are prone to collide, resulting in no cooperation at all.

Opportunistic relaying schemes [2] aim at selecting the "best" relay node, by having relay candidates overhearing the transmission of a RTS from the source and a CTS from the destination. During these transmissions, relay candidates estimate the *Channel State Information* (CSI) from the source and from the destination by overhearing RTS and CTS trans- missions, respectively. The estimated CSI values are then used to compute the end-to-end performance, being the relay candidate providing the best performance factor selected as relay.

Opportunistic relaying as the drawback of requiring knowl- edge about the CSI of all the relay candidates. Hwang et al. [?] introduced an optimized relaying process aiming to minimize the need for CSI estimation, but the proposed scheme always select a relay regardless of the quality of the source-destination channel, and the limitations of relay candidates. Studies about relay limitations have been presented in the literature[3], namely related to battery power of candidate nodes, aiming to select relays to increase network lifetime.

Although most of the related work considers opportunistic relaying, it may lead to packet collision if more than one relay is selected. Collisions may be avoided by using a suitable resource allocation scheme, or by using a relay only when needed, which leads to the need to devise a relay on demand mechanism. On demand relaying may be triggered by a destination of by a relay candidate. In the latter case, a selected relay may estimate the need to use an additional relay, for instance when its channel to the destination is still below the required performance levels. On-demand selection mech- anisms triggered by the destination[1] are characterized by two factors: i) the decision to initiate the selection procedure may depend upon the energy required to receive data; ii) relay candidates may decide not to take part in relaying mechanism. In general, on-demand relay selection works based on the CSI that the destination is able to estimate after receiving a RTS from the source. Based on the CSI, the destination computes the *Packet Error Rate* (PER), being the cooperation used if the PER is above a certain threshold. However, on-demand relay selection relies on RTS and CTS overhearing, leading to an increase of the communication overhead. Moreover, proposed mechanisms do not take advantage of full diversity, because the cooperation occurs only at higher PER, and their complexity may increase in multi-hop scenarios.

All the analyzed approaches try to decrease outage and/or energy expenses. For instance, RelaySpot [5] aims to provide gain in throughput, by having the source selecting a relay that can increase the data rate of the direct link. The source decision is based on a coopTable, which is updated periodically as all nodes broadcast their rate information. The major limitation comes from periodic broadcasting which can affect the network performance.

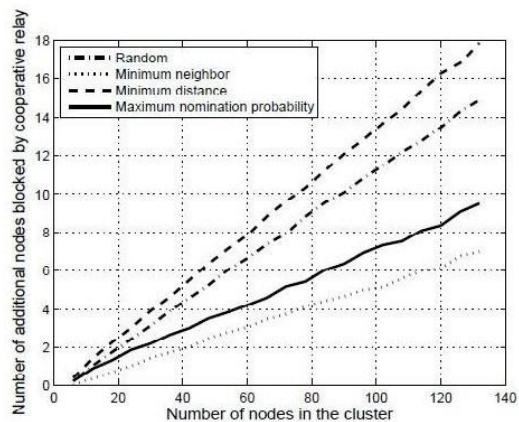

Fig. 2. Spatial efficiency [6]

One potential drawback of cooperative relaying is that coop- erative communication may block additional non-cooperative transmissions (cf. 2). Some studies have been conducted to re- duce the blocking effect of cooperative relaying[6]. A common method is to assign priorities to relay candidates based on their distance to the source and destination. Relays near a source will have a lot in common with the source transmission space, reducing the probability of blocking additional transmissions. However, selecting a relay near a source results in a low spacial diversity.

### III. PROPOSED RELAY SELECTION ALGORITHM

Our 802.11 backward compatible relay selection strategy, called "Relaying on Spot", aims to ensure accurate and fast relay selection, posing minimum overhead and reducing the dependency upon channel estimations, in scenarios where nodes are moving fast. In "Relaying on Spot", relaying decisions take place at potential relay nodes, based on a balanced usage of inter-relay cooperation and opportunistic relaying. Intermediate nodes take the opportunity to relay in the presence of local favorable conditions (e.g. no concurrent traffic) and absence of relaying attempts by any other node. Cooperation between relays is activated as soon as another potential relay heard that the first relay attempt reached the destination in poor condition.

Most of the related work neither analyze the advantages of having more than one simultaneous relay (multiple-relay scenario), nor investigate for how long the selected relay or relays should be kept as the most suitable ones [7]. This is an important question due to the high variability of the wireless channels and the fact that relays may be mobile. The usage of more than one relay is motivated by the unitary complexity ratio of multiple-relay approaches in relation to single-relay approaches. This gives room to investigate a multi-relay selection mechanism able to achieve high diversity, throughput and robustness. In such a scenario (c.f. 3), relays may be single-hop or multi-hop relays and may operate in parallel or in sequence. Additionally, different relays may have different responsibilities.

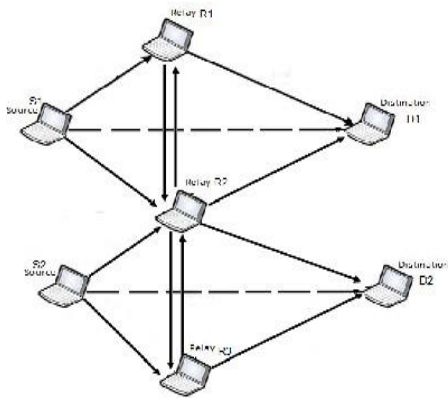

Fig. 3. Concurrent cooperative relaying

For instance, one relay can be active while another can be passive, being their role decided based on their channel conditions. Based on such conditions, the passive relay may be available to be used when needed. Such need may raise when the quality of the active relay channel gets weaker, when the active relay is running out of energy, or when the active relay is moving away [8].

### IV. CONCLUSIONS

In this paper we studied various relay selection techniques recently proposed in literature. Most of the analyzed techniques propose to use channel state information for selecting a relay, while some of them uses distance and performance matrix too. However, few prior art focus on the impact that cooperative relaying have in concurrent transmissions or on the advantage of having cooperation among multiple relays. In this position paper, we argue that the usage of relaying on spot algorithms based on the cooperation among several relays may bring benefits to transmission throughput and network lifetime.

As a future work we will implement this relay selection mechanism using simulator in healthcare scenarios [9][10][11].

REFERENCES

[1] T. Jamal and Z. Haider, "Denial of Service Attack in Cooperative Networks", in Proc. of ArXiv, arXiv: CoRR Vol. arXiv:1810.11070 [cs.NI], Oct. 2018.
[2] T. Jamal and P. Mendes, "Relay Selection Approaches for Wireless Cooperative Networks", in Proc. of IEEE WiMob, Niagara Falls, Canada, Oct. 2010.
[3] T. Jamal, P. Mendes, and A. Zúquete, "Opportunistic Relay Selection for Wireless Cooperative Network", in Proc. of IEEE IFIP NTMS, Istanbul Turkey, May 2012.
[4] T. Jamal and P. Mendes, "Cooperative relaying in user-centric networking under interference conditions", in Proc. of IEEE Communications Magazine, vol. 52, no. 12, pp. 18–24, Dec 2014.
[5] T. Jamal, P. Mendes, and A. Zúquete, "Relayspot: A Framework for Opportunistic Cooperative Relaying", in Proc. of IARIA ACCESS, Luxembourg, June 2011.
[6] S. A. Butt, T. Jamal, and M. Shoaib, "IoT Smart Health Security Threats," in proc. of 19th International Conference on Computational Science and Its Applications (ICCSA), Saint Petersburg, Russia, 2019, pp. 26-31. doi: 10.1109/ICCSA.2019.000-8.
[7] T. Jamal, M Alam, and MM Umair, "Detection and Prevention against RTS Attacks in Wireless LANs", in Proc. of IEEE C-CODE, Islamabad Pakistan, Mar. 2017.
[8] T. Jamal, P. Mendes, and A. Zúquete, "Interference-Aware Opportunistic Relay Selection", In Proc. of ACM CoNEXT student workshop, Tokyo, Japan, Dec. 2011.
[9] T. Jamal and P. Mendes, "Analysis of Hybrid Relaying in Cooperative WLAN", In Proc. of IEEE IFIP Wireless Days (WD), Valencia, Spain, November 2013.
[10] SA Butt and T. Jamal, "A multivariant secure framework for smart mobile health application", in Transactions on Emerging Telecommunications Technologies, Aug. 2019.
[11] M. Asam and T. Jamal, "Security Issues in WBANs", in proc of Arxiv, Volume arXiv:1911.04330 [cs.NI], November 2019.